\documentclass[aps,pra,reprint,amsmath,amssymb,floatfix,superscriptaddress]{revtex4-1}

\usepackage{physics}
\usepackage{color}
\usepackage{graphicx}
\usepackage{times,psfrag}
\usepackage{dsfont}
\usepackage{dcolumn}
\usepackage{bm,bbm}
\usepackage{latexsym,amsmath,amssymb,bm,euscript}
\usepackage[normalem]{ulem}
\usepackage[hidelinks]{hyperref}
\usepackage{microtype}
\usepackage{verbatim} 

\newcommand{\WSe}{WSe$_{2}$}
\newcommand{\MoSe}{MoSe$_{2}$}
\newcommand{\WS}{WS$_{2}$}
\newcommand{\MoS}{MoS$_{2}$}

\newcommand{\SiO}{SiO$_{2}$}
\newcommand{\SiN}{Si$_{3}$N$_{4}$}

\begin{document}
\title{%
Biaxial strain tuning of excitons in monolayer MoSe$_2$ by high-temperature physical vapor deposition
}
%
\author{S. Patel}
\affiliation{Department of Physics, University of Arkansas, AR 72701, USA}
\author{T. Faltermeier}
\affiliation{Department of Physics, Montana State University, Bozeman, MT 59717, USA}
\author{S. Puri}
\affiliation{Department of Physics, University of Arkansas, AR 72701, USA}
\author{R. Rodriguez}
\affiliation{Department of Physics, University of Arkansas, AR 72701, USA}
\author{K. Reynolds}
\affiliation{Department of Physics, University of Arkansas, AR 72701, USA}
\author{S. Davari}
\affiliation{Department of Physics, University of Arkansas, AR 72701, USA}
\author{H. O. H. Churchill}
\affiliation{Department of Physics, University of Arkansas, AR 72701, USA}
\author{N. J. Borys}
\affiliation{Department of Physics, Montana State University, Bozeman, MT 59717, USA}
\author{H. Nakamura}
\email{hnakamur@uark.edu}
\affiliation{Department of Physics, University of Arkansas, AR 72701, USA}

\begin{abstract}
We present strain tuning of excitonic emission in monolayer MoSe$_2$ by using a high-temperature physical vapor deposition (PVD). The use of two amorphous substrates, \SiN\ and \SiO, provides two setpoints to induce distinct amounts of \textit{biaxial} tensile strain determined by a thermal expansion mismatch between the monolayer and the substrate. The tuning rate of the $A$-exciton transition energy is found to be 103 meV/\% by photoluminescence (PL), which represents the highest value realized by biaxial strain in transition metal dichalcogenides. The biaxial nature of the tensile strain is confirmed by polarization-resolved second harmonic generation, which reveals unperturbed in-plane three-fold symmetry of the monolayer. Furthermore, a softening of $A_\mathrm{1g}$ out-of-plane lattice vibration is identified in the Raman spectroscopy, which is known to be insignificant for uniaxial strain. Concomitantly, PL mapping of our PVD monolayers demonstrates (i) larger strain occurs in the interior of the mono-domain islands compared to the edges and (ii) the absence of island-size dependence in the magnitude of induced strain. Our results demonstrate an effective path towards strain engineering of excitons by using growth substrates, which holds great promise as a building block for future optoelectronic applications.

\end{abstract}

\maketitle

\section{Introduction\label{intro}}
Since the indirect-to-direct band gap transition was found in the monolayer limit of transition metal dichalcogenides (TMDs) \cite{mak2010,splendiani2010emerging}, excitonic states and resulting optoelectronic properties in atomically thin TMDs have attracted significant interest \cite{mak2016photonics,wang2018colloquium}. The excitonic binding energy in monolayer TMDs are giant, of the order of 100-500 meV, making them suitable for studying various aspects of excitonic physics \cite{rivera2018interlayer,ma2021strongly,jia2022evidence}.

One of the appealing features of 2D excitons is that they are sensitive to external perturbations, including changes in the dielectric environment \cite{stier2016exciton,raja2017coulomb} as well as strain \cite{conley2013bandgap,he2013experimental,liu2014strain,island2016precise,ji2016strain,lloyd2016band,
horzum2013phonon,chae2017substrate,plechinger2015control,frisenda2017biaxial,hu2018mapping,
cun2019wafer,luo2020investigation,covre2022revealing,lei2021size,seravalli2023built}. For example, in experiments where mechanical strain is applied to monolayer TMDs, changes in excitonic emission energies as large as $\sim$100-200 meV have been reported, which corresponded to maximum strain levels of 1 to 2\% \cite{roldan2015strain}. However, these mechanically-induced strains typically induce uniaxial strain rather than uniform biaxial strain, albeit with a few notable exceptions \cite{lloyd2016band}. Theories predict distinct impact of strain on electronic and phonon dispersions for uniaxial and biaxial strain cases \cite{chang2013orbital}, but experimental studies on biaxial strain are relatively limited compared to those for uniaxial strain.

A natural route to induce biaxial strain is through a substrate on which a monolayer is grown or transferred.
For example, when the substrate has the same in-plane lattice symmetry as the film to be grown, strain is caused by the difference in the in-plane (super)lattice spacing of the substrate and the monolayer film \cite{nakamura2020spin}. Strain induced in 2D materials on an amorphous substrate by a mismatch in thermal expansion (TE) has also been studied, by using methods such as local heating, examining a non-uniform strain caused during the chemical vapor deposition (CVD), or during the cooling process after the growth. For example, Ahn \textit{et al.} \cite{ahn2017strain} studied biaxial strain in CVD-grown monolayer \WSe\ and reported a redshift of the exciton transition energy of $\sim$75meV with $\sim$0.9\% tensile strain (i.e.,83 meV/\%) on \SiO\ grown at 900$^\circ$C. Cun \textit{et al}. \cite{cun2019wafer} reported a $\sim$20 meV red shift in the PL of \MoS\ on \SiO\ grown at 850$^\circ$C. On the other hand, Lei \textit{et al}. \cite{lei2021size} and Severalli \textit{et al}. \cite{seravalli2023built} showed locally inhomogeneous strain in CVD-grown \MoS\, in which the corner and edges of triangular islands showed greater tensile strain than the interior region. In contrast, Luo \textit{et al.} showed, also by using a CVD-grown monolayer \MoS, that strain was larger in the interior region than in the edges, exhibited by a lower exciton transition energy (red shift) in the interior of a monolayer island. \cite{luo2020investigation}.

From these earlier studies, we can deduce that (i) TE mismatch could transfer sizable biaxial strain from substrates to monolayer van der Waal (vdW) materials, (ii) experiment and theory both show a redshift of the exciton transition energy via an application of biaxial tensile strain, (iii) uniform biaxial strain through TE mismatch was observed in \WSe\ \cite{ahn2017strain}, but noticeable inhomogeneity in strain using the same approach was identified in \MoS\ \cite{lei2021size,seravalli2023built,luo2020investigation}, and (iv) the spatial distribution of strain within a monodomain island can be qualitatively different even for the same type of TMD \cite{lei2021size,seravalli2023built,luo2020investigation}. We point out that all of these studies were carried out using CVD-grown monolayers, and the final point may be impacted by different growth conditions. It is also worth noting that spatially-resolved studies of \textit{biaxial} strain have mostly been limited to monolayer \MoS. In the light of interest in strain-engineering of 2D materials including spatially inhomogeneous structures \cite{roldan2015strain,naumis2023mechanical}, it is highly desirable to advance our understanding of the effects of biaxial strain induced by TE mismatch in other monolayer materials.
\begin{figure*}[!bt]
\includegraphics[width=14cm,clip]{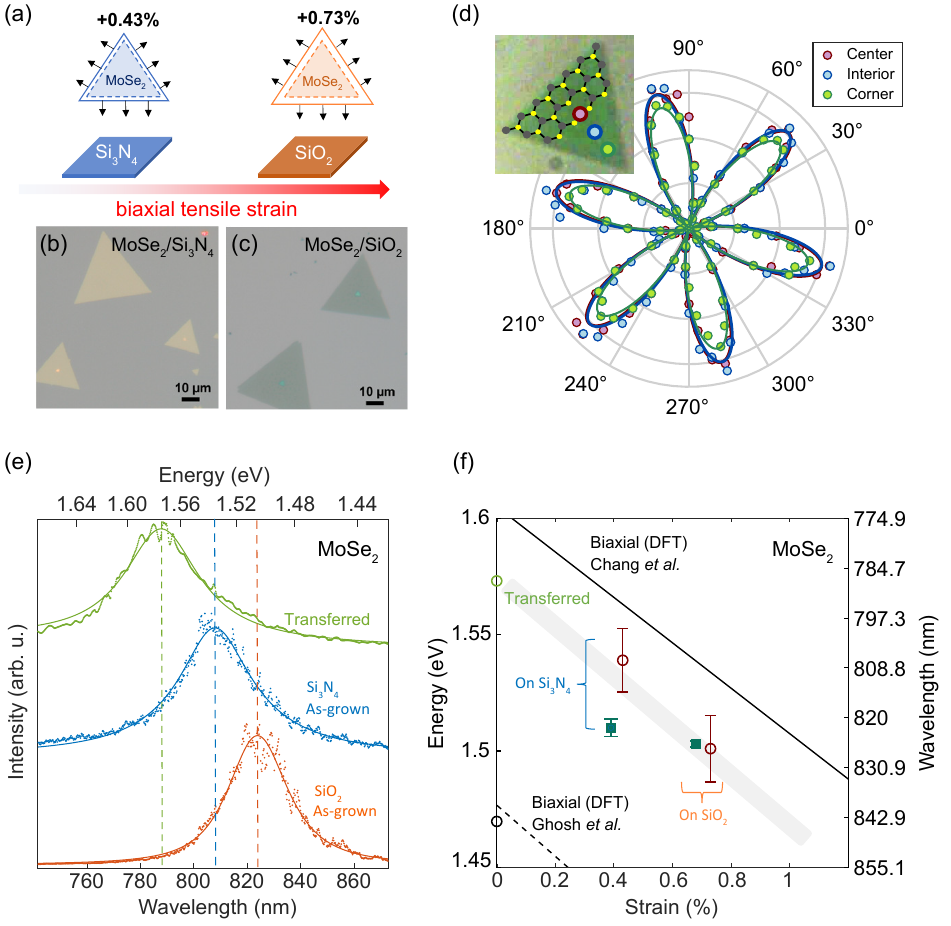}
\caption{%
Biaxial strain tuning of monolayer \MoSe. (a) Schematic diagram showing the expected tensile strain induced in monolayer \MoSe\ 
through TE mismatch from the substrates. Optical microscope images of (b) \MoSe\ on \SiN\ and (c) \MoSe\ on \SiO\ grown by high-temperature PVD. (d) Polarization resolved SHG taken at three different locations on the monolayer island grown on \SiO. The location of laser spots are indicated in the inset. The SHG was taken with the analyzer in the parallel polarization configuration with respect to the fundamental wavelength. Solid lines are a fit to $\sin^2(3\theta-\phi)$, where $\theta$ shows the angle of fundamental polarization from the laboratory $x$ (horizontal) axis, while $\phi$ denotes the angular offset of an armchair direction of the island with respect to the $x$ axis. (e) Representative PL data taken from the interior region of as-grown \MoSe\ islands on \SiN\ and \SiO\ at 1130 $^\circ$C, and that from an island transferred on another \SiO\ substrate. Experimental data is plotted as dots, on which Lorentzian fits are overlayed (solid lines). Vertical dashed lines shows the peak position derived from the fit. (f) The $A$ exciton transition energy as a function of biaxial tensile strain. Black solid and dashed lines indicate the results from DFT calculations \cite{ghosh2013equibiaxial,chang2013orbital}. Brown open circles (green filled squares) represent A-exciton energy for \MoSe\ grown at 1130$^\circ$C (1075$^\circ$C), while the green open circle represents that for transferred \MoSe. PL from 15-20 triangular islands were used to obtain averaged exciton energy for each data point.
}
\label{fig1}
\end{figure*}

In this paper, we study the biaxial strain tuning of excitons in monolayer \MoSe\ through TE mismatch, by using amorphous \SiO\ and \SiN\ substrates to create a controllable strain. The significance of our approach can be summarized in three points. First, we adopt a high growth temperature ($\sim$1100$^\circ$C) that increases $\Delta T$, the difference between the growth temperatures and room temperature, thereby enhancing the expected amount of strain due to TE mismatch. Second, we utilize physical vapor deposition (PVD) that uses only a single compound as a precursor, reducing complexities that may arise from the use of multiple precursors in CVD. Third, the use of amorphous  \SiO\ and \SiN\ substrates eliminates epitaxial matching effects, allowing us to focus on a scenario where TE mismatch dominates. By adopting this approach, we show that the substrate-induced strain can be systematically tuned in as-grown monolayer TMDs, demonstrating a large strain tuning ratio of 103 meV/\% for the $A$ exciton in \MoSe\ monolayers. Furthermore, $\mu$PL imaging and spectroscopy demonstrates greater tensile strain in the interior of island than in the edges, as well as the lack of island size dependence of strain. Polarization-resolved second harmonic generation provides further evidence for the homogeneous nature of biaxial strain in the interior of triangular islands. Our results demonstrate a simple path towards uniform and controllable strain in as-growth monolayers, which could be highly beneficial for fundamental studies of excitonic physics as well as for future optoelectronic devices.

\section{Experiment}

Monolayer \MoSe\ films were grown using a home-built PVD system at a growth temperature of 1075$^\circ$C or 1130$^\circ$C using pure \MoSe\ powder as a source material. Note that these growth temperatures are much higher than those typically used for the CVD growth of TMDs (800-900$^\circ$C). The PVD system is equipped with two mass flow controllers installed on both sides of a quartz tube (diameter: 1 inch) that regulate the Ar flow to prevent unwanted deposition. Inhibiting unwanted deposition is achieved by flowing Ar from the substrate side towards the powder side outside of the intended growth time. Following a growth time of 5-10 seconds, we rapidly quenched the temperature by sliding the furnace away from the location of the powder and the substrate. Substrates were silicon with custom grown oxide and nitride layers of 90 nm (\SiO) and 70nm (\SiN) in thickness, optimized to enhance the optical contrast of monolayers \cite{rubio2015enhanced,zhang2015measuring}. We transferred as-grown monolayer islands using a nail polish technique \cite{haley2021heated} to a fresh \SiO\ substrate to release strain, which we used as an unstrained reference. Atomic force microscopy (AFM) was used to check the thickness of the TMD crystallites. Single point photoluminescence (PL) and Raman measurements were carried out using a home built system equipped with a 532 nm excitation laser. Objective lenses with a 50x (NA=0.55) and 100x (NA=0.8) magnifications were used to focus the light on the sample, and the laser power on the sample was below 2 mW. All the Raman/PL measurements were performed in air at room temperature. We used two narrow-line OD4 Bragg type filters (Optigrate) to reject the laser line, which allowed us to measure Raman shifts down to $\sim$20 cm$^{-1}$ for both Stokes and Anti-Stokes peaks. Silicon Raman peaks from the substrate were used to calibrate the Raman shifts. The diffraction limited $\mu$PL imaging and spectroscopy was performed on a HORIBA Trios microscope using a 100x (NA=0.8) objective and a 532 nm excitation laser (Coherent). The back-reflected laser light was filtered from the PL emission from the sample by long-pass filters (Semrock). The isolated PL emission was detected by an EMCCD attached to a spectrometer (Andor). Second harmonic generation (SHG) was measured by using Ti:sapphire fs oscillator (Tsunami, Spectra-Physics) as an excitation laser, which delivered a temporal pulse width of $\sim$200 fs with $\sim$ 1 mW at the position of the sample, and reflected SHG was observed through filters that rejected the fundamental beam. A spectrometer equipped with a thermoelectrically-cooled CCD camera was used to detect a second harmonic signal.

\section{Results and discussion}\label{res}

\subsection{Substrate-dependent biaxial strain}\label{pl}

The morphology of grown monolayer TMDs was measured by optical microscopy and AFM. As can be seen from the optical images (Fig. 1-b,c), triangular monolayer islands of 10-60$\mu$m in size were obtained by the high temperature PVD growth. The AFM showed the thickness of triangular island to be approximately 0.6 nm, consistent with the expected height of monolayers. Some islands show indication of a thicker overgrowth at the center, which can be seen by a different contrast in optical images. Monolayer islands with more complex shapes such as polygons are also occasionally observed. In this study, we focus on the data collected from a triangular monolayer. Except for the spatially-resolved PL, which we describe later, the PL or Raman data were taken from an interior region of each triangular monolayer, which we define to be between the edge and the center of each triangular island.

A representative PL spectrum of monolayer \MoSe\ on the two substrates (\SiO\ and \SiN) taken from such regions are shown in Fig. 1e. What is immediately noticeable is that the $A$ exciton peak resdshifts for monolayers on \SiO\ compared to \SiN. We attribute this red shift to biaxial substrate strain due to thermal expansion mismatch between the monolayer and the substrate, which is larger for \SiO\ than for \SiN. The uniform, biaxial nature of strain is evident from the 60$^\circ$ periodic pattern (flower pattern) in the polarization resolved SHG [Fig.1(d)], which retains a three-fold rotation symmetry about the surface normal ($z$ axis). This is in contrast with the uniaxial strain case, in which such a rotational symmetry is absent in polarization-resolved SHG \cite{mennel2018optical}. By SHG analysis, we also note that the edge of our triangular islands were confirmed to be always along the zigzag direction of the \MoSe\ lattice [inset of Fig.1(d)]. Further details on the biaxial-strain tuning of SHG is reported elsewhere \cite{puri2024substrate}.

The strain induced in the TMD film can be estimated by the following equation,
\begin{equation}
\mathrm{Strain(\%)} = \left(\alpha_{\mathrm{TMD}}-\alpha_\mathrm{sub}\right) \cdot \Delta T \cdot 100,
\end{equation}
where $\alpha_{\mathrm{TMD}}$ is the thermal expansion coefficient of \MoSe, $\alpha_\mathrm{sub}$ is that of \SiN\ or \SiO, $\Delta T = T_\mathrm{growth}-T_\mathrm{room}$ in which $T_\mathrm{growth}$ and $T_\mathrm{room}$ denote the growth temperature and room temperature, respectively. This is an approximate formula, which uses average $\alpha$ values for the temperature range of interest. 
We used the following values for thermal expansion coefficients: $\alpha_{\mathrm{MoSe_2}}=7.1\times10^{-6}$ \cite{el1976thermal}; 
$\alpha_\mathrm{SiO_2}=5.0\times10^{-7}$ \cite{roy1989very}; $\alpha_\mathrm{Si_3N_4}=3.5\times10^{-6}$ \cite{sinha1978thermal}. By using $T_\mathrm{room}=25^\circ$C and $T_\mathrm{growth}=1130^\circ$C (1075$^\circ$C), we estimate a tensile strain of +0.43\% (+0.41\%) on \SiN\ and +0.73\% (+0.69\%) on \SiO. This mechanism thus can induce a sizable tensile strain in the TMD films when the film is grown at high temperatures, particularly on \SiO\ substrates.

The biaxial-strain tuning of the exciton transition energy is summarized in Fig. 1(f). We obtained a tuning rate of 103 meV/\% from the linear fit of the experimental data obtained from as-grown and transferred monolayers. We point out that Frisenda \textit{et al.} experimentally derived a tuning rate of 33 meV/\% by heating a plastic substrate, while Covre \textit{et al.} estimated that of 54 meV/\% by studying a bubble structure, both in monolayer \MoSe. Thus, the tuning rate of 103 meV/\% obtained in the present work is the highest value experimentally observed in biaxially-strained \MoSe\ monolayers. In Fig. 1(f), we also plot the results of biaxial strain tuning from first-principles DFT calculations \cite{chang2013orbital,ghosh2013equibiaxial}. The experimental strain tuning rate is in good agreement with the ones derived by theories [97.5 meV/\% and 110 meV/\%, see Fig. 1(f)].

\begin{figure*}[!bt]
\includegraphics[width=14cm,clip]{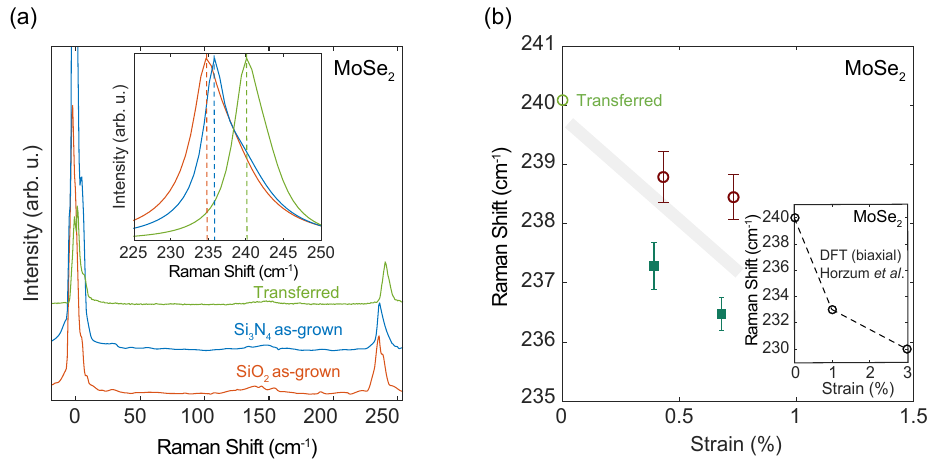}
\caption{%
Biaxial tensile strain as probed by Raman spectroscopy. (a)  Representative Raman data taken from the interior region of as-grown \MoSe\ islands on \SiN\ and \SiO\ and that from an island transferred on a fresh \SiO\ substrate. The inset shows the $A_\mathrm{1g}$ phonon peak for monolayer on each substrate. (b) Raman shift of the $A_\mathrm{1g}$ phonon as a function of biaxial tensile strain for \MoSe\ grown at 1075$^\circ$C (filled squares) and 1130$^\circ$C (open circles), and for \MoSe\ transferred on a fresh \SiO. Raman data from several monolayer islands were averaged to obtain each data point. A grey line shows a linear fit to the data. The inset shows the result from DFT calculation \cite{horzum2013phonon}, where the dashed line is used to connect data points derived from the calculation.
}
\label{fig2}
\end{figure*}
   
The occurrence of biaxial strain can be further confirmed by Raman spectroscopy. Under tensile strain, selected phonon modes of monolayer TMDs soften and exhibit smaller Stokes shifts, providing an additional spectroscopic signature of strain. Importantly, phonon modes respond distinctly to uniaxial and biaxial strains. Specifically, out-of-plane A$_\mathrm{1g}$ phonons in monolayer \MoS\ was found to soften with the application of biaxial strain \cite{lloyd2016band}, while no softening was observed with uniaxial strain \cite{conley2013bandgap}. We focus on A$_\mathrm{1g}$ phonons with Raman shifts of 240.5 cm$^{-1}$ \cite{tonndorf2013pl} in an exfoliated monolayer \MoSe\ at room temperature. We first note that the Raman shift obtained from a transferred monolayer is 240.1 cm$^{-1}$ [Fig. 2(b)], which is very close to the the literature value of an exfoliated one, consistent with the expectation that the strain is released by transferring. The $A_\mathrm{1g}$ Raman modes in as-grown monolayers, on the other hand, show substantial a red-shift depending on the substrate [Fig. 2(a)]. This is further evidence that the substrate-induced tensile strain is biaxial in nature. We plot the Raman shift values with respect to estimated biaxial strain in Fig. 2(b). A linear fit to the data points yields a tuning rate of$\Delta A_{1g} = 3.35$ cm$^{-1}$/\%. The inset of Fig.2(b) also shows a result from a density functional theory (DFT), which found a tuning rate of $\Delta A_{1g} \sim$7 cm$^{-1}$/\% for a strain between 0-1\%, beyond which the $\Delta A_{1g}$ becomes much smaller ($\sim$1 cm$^{-1}$/\%) \cite{horzum2013phonon}. The theory suggests a nonlinear relationship between strain and the redshift of $A_\mathrm{1g}$ Raman mode. Further study is needed for quantitative analysis of the Raman red-shift in PVD-grown \MoSe\ monolayer.

\begin{figure*}[!bt]
\includegraphics[width=17cm,clip]{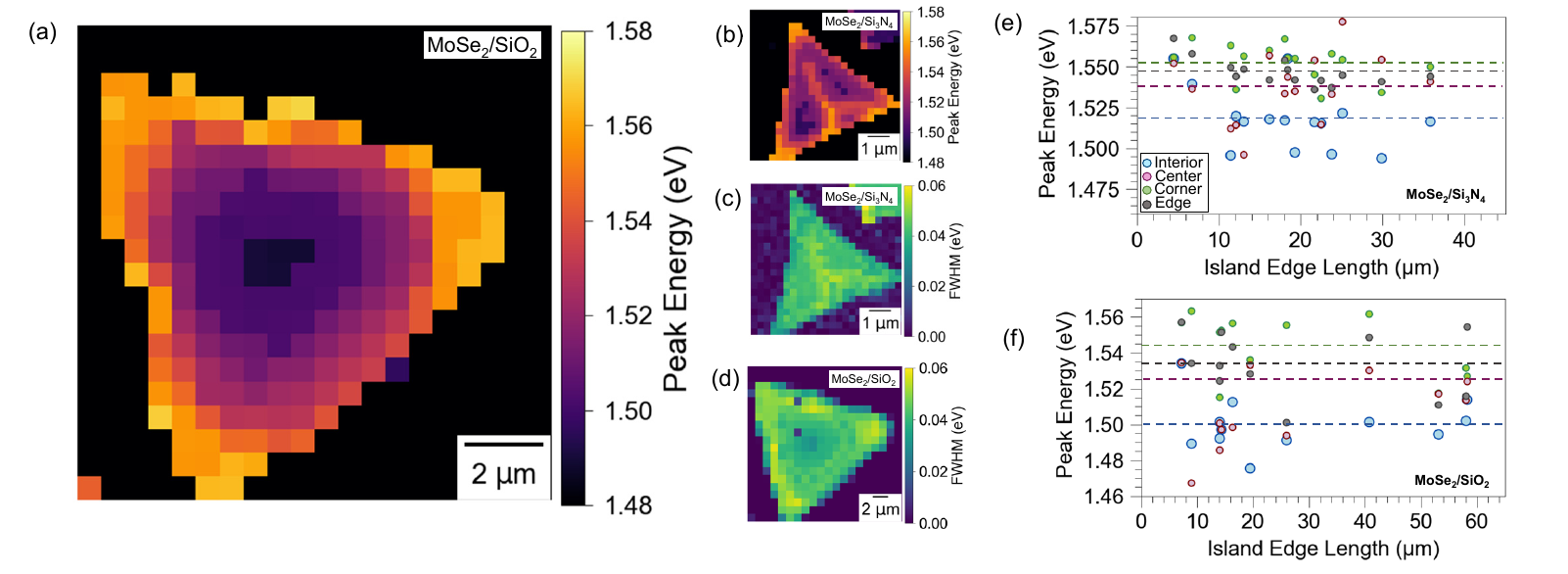}
\caption{%
Spatial mapping of strain in monolayer \MoSe. Spatially resolved PL mapping on (a) monolayer \MoSe\ on \SiO\ and (b) on \SiN, showing the $A$ exciton peak energy in each pixel position with a color scale on the right. The position of three circles and the line in (a) defines corner, interior, center, and edge regions, from which we obtain representative values of exciton energy after averaging all the values for pixels that overlap with these symbols. Full width at half maximum (FWHM) of $A$ exciton peaks are shown for (c)  \MoSe\ on \SiO\ and (d) \MoSe\ on \SiN. Island size dependence of $A$ exciton peak energies, taken at four different locations in each island as defined above, are presented in (e) for \MoSe\ on \SiN\ and (f) for \MoSe\ on \SiO. Dashed horizontal lines denote the average peak energy derived for each position, with a color matching those used for the symbols. The edge region consistently shows higher energy than other positions, both for \SiO\ and \SiN\ substrates, indicating reduced tensile strain than other regions.
}
\label{fig3}
\end{figure*}

\begin{figure}[!bt]
\includegraphics[width=\linewidth,clip]{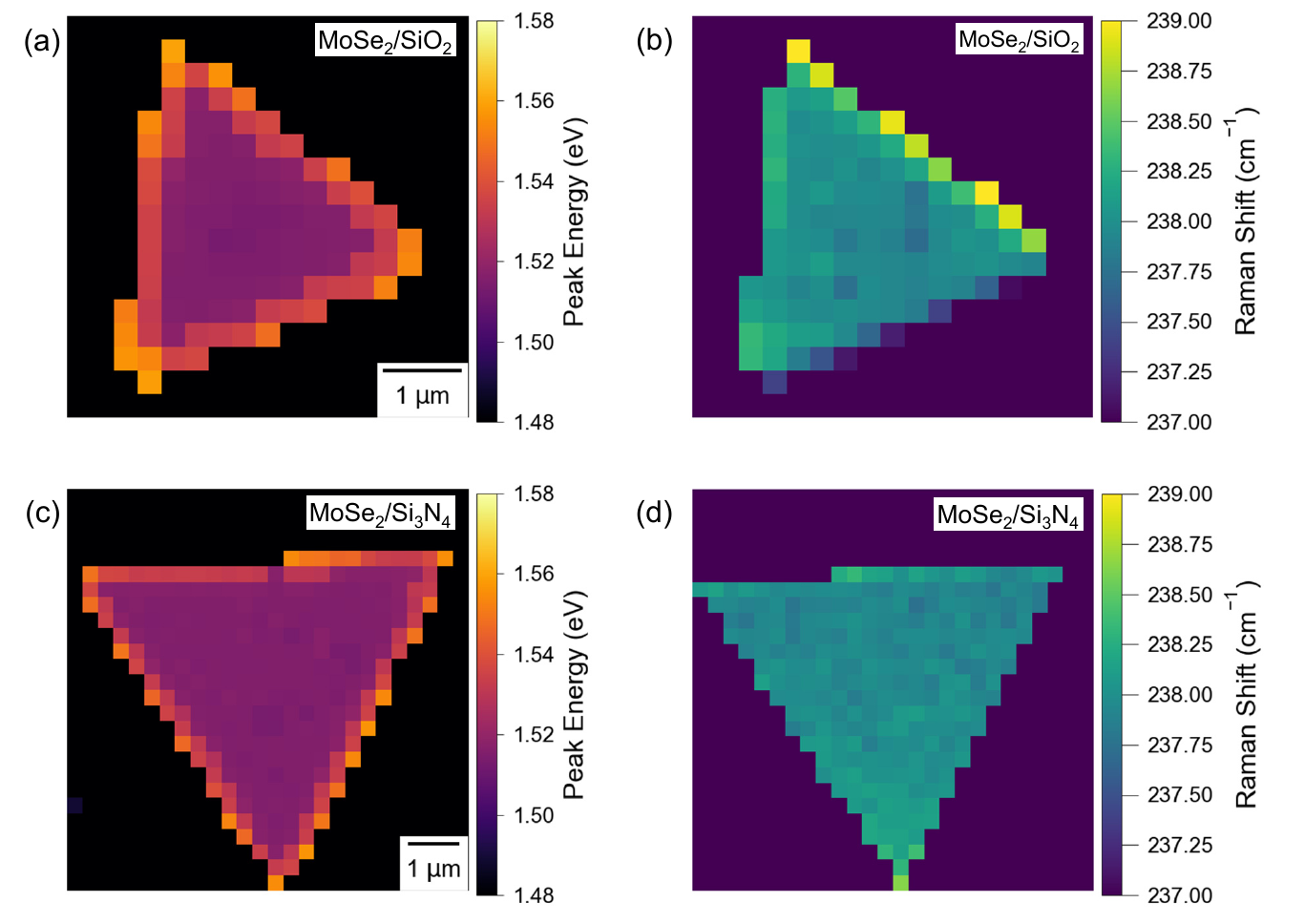}
\caption{%
Comparison of spatially resolved PL [(a),(c)] and Raman [(b),(d)] mapping of \MoSe\ monolayers on the two substrates. The data from the same monolayer are presented in (a),(b), as well as in (c),(d).
}
\label{fig4}
\end{figure}

\subsection{Spatial mapping of biaxial strain}\label{pl}

Spatially-resolved $\mu$PL imaging and spectroscopy was performed on monolayers grown at 1130$^\circ$C  to further elucidate the nature of the biaxial strain and its distribution over single crystallites. 
A representative map of the peak energy of the PL from a monolayer \MoSe\ island on \SiO [Fig. 3(a)] clearly shows that the exciton transition energy red-shifts by 10-30 meV in the interior of the island compared to the edge regions. This trend was common to all of the monolayer \MoSe\ islands we inspected. Note also that the full-width at half maximum (FWHM) of the exciton peak (40-60 meV at room temperature) shows some correlation with the observed red-shift in the exciton transition energy. Interior regions with lower exciton energies correspond to smaller FWHMs. Interestingly, this trend is \textit{opposite} from what was observed in CVD-grown \MoS\ monolayers in Ref. \cite{lei2021size,seravalli2023built}, where larger tensile strain was found at the corners and edges, which in turn led to a larger redshift of the exciton transition energy in those regions compared to the interior. In contrast, near field PL of monolayer \MoS\ in Ref. \cite{bao2015visualizing} found that the interior region has higher energy than a nanoscale edge region, which was attributed to increased disorder at the edges. This is opposite from what we observe here due to strain which we are probing on larger length scales. At these larger length scales, our spatial strain distribution qualitatively matches those of CVD-grown \MoS\ and \WS\ in Ref. \cite{luo2020investigation,kastl2017important}, which showed that the interior regions exhibited lower exciton transition energy (i.e., a larger red shift) than the perimeters. Comparing the spatial dependence of strain in as-grown \MoS\ monolayers indicate that strain-formation from TE mismatch may depend sensitively on the specific growth conditions used.

To shed light on a possible strain formation or release mechanism, we show exciton transition energies extracted from representative positions in each island with respect to the size of the island [Fig. 3(e,f)]. If the tensile strain, or its release, is triggered by any mechanism that depends on the size of island, such plots would show a correlation between the exciton energies and the size of the island. In the plots in Fig. 3(e,f), however, we do not see any clear trend in peak energy with respect to the size of the island. Instead, these plots reconfirm that the edge of the island consistently exhibits a higher peak energy compared to the interior region of the same island, for both \SiO\ and \SiN\ substrates, without any noticeable trend that depends on its size. We also note that when only the edge regions are compared between the two substrates, the photoluminescence from the edge of \MoSe\ on \SiO\ is red shifted compared to that on \SiN. This trend indicates that the tensile strain exists in the edge region, but is partially released compared to the interior of each monodomain island. Besides the reduced red shift in PL, we note that the intensity of SHG is also reduced near the perimeter of an island [Fig. 1(d)]. Combined spatially resolved Raman PL imaging of monolayers on each substrate show signatures of the strain relaxation at the edges (Fig. 4). For both \SiO\ and \SiN\ substrates, the PL at the edges is shifted to higher energies by ~30-50 meV. Similarly, there are subtle signatures that the Raman modes are also higher in energy in these regions as well, but the observed shifts are smaller than what would be anticipated from the PL spectroscopy of the excitons. These differences -- which motivate further exploration with, for instance, nano-optical techniques -- could reflect strain-induced variations in carrier density or the exciton diffusion effects which would have greater effects on the observed PL energies than in the Raman spectroscopy.
   
\section{Conclusions\label{sum}}
In summary, we have demonstrated biaxial strain tuning of monolayer \MoSe\ by using high-temperature PVD. This method provides an effective way to transfer biaxial tensile strain to the monolayer through a TE mismatch during the growth, which enabled a large tuning rate of 103 meV/\% for the $A$ exciton transition energy. The polarization-resolved SHG showed three-fold in-plane symmetry of the monolayer to be intact without indication of anisotropy, while Raman spectroscopy demonstrated a softening of the out-of-plane $A_\mathrm{1g}$ phonon mode, both of which are consistent with the biaxial nature of strain. Spatially resolved $\mu$PL imaging and spectroscopy shows that the edge of the monolayer islands consistently has higher energy excitons compared to the interior region. The exciton energies do not exhibit a clear correlation with the size of the grown islands. Biaxial strain tuning by high-temperature PVD could be applied to other monolayer TMD and 2D materials, and should provide a powerful tool for the strain-engineering of atomically-thin materials.

\begin{acknowledgments}

The work is supported by the Office of the Secretary of Defense for Research and Engineering under Award No. FA9550-23-1-0500 and the MonArk NSF Quantum Foundry from the National Science Foundation under Award No. DMR-1906383.

\end{acknowledgments}

\bibliographystyle{apsrev4-1} 

\end{document}